# Quartic Lattice Interactions, Soliton-like Excitations and Electron Pairing in One-Dimensional Anharmonic Crystals

M. G. VELARDE[1], L. BRIZHIK[1,2], A. P. CHETVERIKOV[1,3],

L. CRUZEIRO[1,4], W. EBELING[1,5], G. RÖPKE[1,6]

[1]Instituto Pluridisciplinar, Universidad Complutense, Paseo Juan XXIII, 1, Madrid 28040, Spain

[2]Bogolyubov Institute for Theoretical Physics, Metrolohichna Str., 14b, Kyiv 03680, Ukraine

[3]Faculty of Physics, Chernyshevsky State University, Astrakhanskaya Str., 83, Saratov 410012, Russia

[4]CCMAR and FCT, Universidade do Algarve, Portugal, Campus de Gambelas, Faro 8005-139, Portugal

[5]Institut für Physik, Humboldt Universität, Newtonstrasse 15, Berlin 12489, Germany

[6]Institut für Physik, Universität Rostock, Universitätsplatz, 3, Rostock 18051, Germany

**ABSTRACT:** We show that two added, excess electrons with opposite spins in one-dimensional crystal lattices with *quartic* anharmonicity may form a *bisolectron*, which is a localized bound state of the paired electrons to a soliton-like lattice deformation. It is also shown that when the Coulomb repulsion is included, the wave function of the bisolectron has two maxima, and such a state is stable in lattices with strong enough electron-(phonon/soliton) lattice coupling. Furthermore the energy of the bisolectron is shown to be lower than the energy of the state with two separate, independent electrons, as even with account of the Coulomb repulsion the bisolectron binding energy is positive.

## 1. Introduction

In a previous publication [1] we have considered soliton-mediated electron pairing in *anharmonic* lattices endowed with *cubic* inter-particle interaction. The latter is of mathematical interest as it underlies as a *dynamical system* the soliton-bearing Boussinesq-Korteweg-de Vries equation governing wave propagation along the lattice in the continuum approximation [2-4]. Noteworthy is that due to the electron-lattice interaction, the soliton arising from the cubic interaction is able to trap an electron thus leading to a *solectron* [5-7] in a way that generalizes the *polaron* concept long ago proposed by Landau and Pekar to describe electron self-trapping leading to a *dressed* electron (or, more generally, a quasiparticle) [8-11]. In [1] it was shown that when two *excess* electrons are added the corresponding electron-lattice interaction also facilitates electron pairing leading to a *bisolectron*, i.e., the localized bound state of the paired electrons and a soliton-like lattice deformation. Here we study this problem for a lattice with the more physically appealing *quartic* particle interaction for which no soliton-bearing equation is known to exist.

In Section 2 we pose the problem, introduce the Hamiltonian and the evolution equations of the system. Section 3 is devoted to the search of localized, soliton-like traveling solutions. Then in Section 4 we asses the role of repulsive Coulomb interaction between the two added, excess electrons, satisfying Pauli's exclusion principle. Finally in Section 5 we briefly recall the major results found.

### 2. Hamiltonian and evolution equations

We consider a model one-dimensional, infinitely long crystal lattice with electrons and lattice particles of equal masses *M*, describable by the following Hamiltonian [10,11]:

$$\hat{H} = \hat{H}_{el} + \hat{H}_{lat} + \hat{H}_{int}, \tag{2.1}$$

with

$$\hat{H}_{el} = \sum_{n,s}\left[E_0 \hat{B}^+_{n,s}\hat{B}_{n,s} - J\hat{B}^+_{n,s}\left(\hat{B}_{n+1,s}+\hat{B}_{n-1,s}\right)\right], \tag{2.2}$$

$$\hat{H}_{lat} = \sum_n\left[\frac{\hat{p}_n^2}{2M}+\hat{U}\right], \tag{2.3}$$

$$\hat{H}_{int} = \chi\sum_{n,s}\left(\hat{\beta}_{n+1}-\hat{\beta}_{n-1}\right)\hat{B}^+_{n,s}\hat{B}_{n,s}. \tag{2.4}$$

Note that we have two added excess electrons with opposite spins, $s=\uparrow,\downarrow$. The quantity $E_0$ denotes the on-site electron energy, $J$ denotes the electron exchange interaction energy, $\hat{B}^+_{n,s}$, $\hat{B}_{n,s}$ are, respectively, creation and annihilation operators of an electron with spin $s$ on the lattice site $n$, $\hat{\beta}_n$ is the displacement operator of the *n*-th atom from its equilibrium position, $\hat{p}_n$ is the operator of the canonically conjugated momentum, $\chi$ accounts for the electron-lattice interaction strength, and $\hat{U}=\hat{U}(\{\hat{\beta}_n\})$ is the operator of the potential energy of the lattice, yet to be specified. Only nearest neighbor interactions will be considered and therefore the lattice potential energy depends on their relative displacements:

$$U(\{\beta_n\}) = U(\beta_n - \beta_{n+1}) \equiv U(\rho_n), \tag{2.5}$$

where $\rho_n \equiv \beta_n - \beta_{n+1}$. We take into account only the *longitudinal* displacements of atoms from their initial equilibrium positions (the acoustic mode) and consider that the dependence of the on-site electron energy on lattice atom displacements is much stronger than that of the exchange interaction energy. Further, the electron-lattice interaction is taken strong enough so that the electron self-trapping leads to a bound state whose length extends over several lattice sites. The Coulomb repulsion between the electrons will be included later on.

For two electrons with opposite spins the *antisymmetry* of the two-electron spin function permits representing the two-electron spatial wave-function as a symmetrized product of one-electron wave-functions, $\Psi_{j,n}$, *j=1,2*. The evolution equations for $\Psi_{j,n}$ and for the lattice deformation, $\beta_n$, follow from the variational minimization condition of the Hamiltonian functional obtained from Hamiltonian (2.1). This functional depends on the canonical variables In the continuum approximation, defining $z = na$, with $a$ being the initial equilibrium lattice inter-particle spacing we have:

$$i\hbar \frac{\partial \Psi_j(z,t)}{\partial t} + Ja^2 \frac{\partial^2 \Psi_j(z,t)}{\partial z^2} - \chi a^2 \frac{\partial \beta(z,t)}{\partial t} \Psi_j(z,t) = 0, \qquad (2.6)$$

$$\frac{\partial^2 \beta(z,t)}{\partial t^2} - V_{ac}^2 \frac{\partial^2 U(z,t)}{\partial \rho^2} \frac{\partial^2 \beta(z,t)}{\partial z^2} = \frac{\chi a}{M} \frac{\partial}{\partial z}\left(|\Psi_1(z,t)|^2 + |\Psi_2(z,t)^2|\right), \qquad (2.7)$$

where

$$\rho(z,t) = -a \frac{\partial \beta(z,t)}{\partial z}, \qquad (2.8)$$

Here $V_{ac}^2 = a^2 w/M$ is the sound velocity in the lattice with $w$ being the lattice elasticity constant.

For the lattice inter-particle interaction, $U(\rho)$, we take:

$$U(\rho) = \frac{1}{2}\rho^2 + \frac{\gamma}{4}\rho^4. \qquad (2.9)$$

Note that here we are only interested in the effect of the quartic curvature of the potential relative to the harmonic case and not on the possible influence of a double-well potential.

The one-electron wave-functions in the system of coupled nonlinear equations (2.6)-(2.7) satisfy the normalization condition

$$\frac{1}{a}\int_{-\infty}^{\infty} |\Psi_j(z,t)|^2 dz = 1 \quad (j=1, 2), \qquad (2.10)$$

and for the stationary states can be written in the form:

$$\Psi_j(z,t) = \Phi_j(\xi)\exp\left\{\frac{i}{\hbar}\left[mVz - E_j t - \frac{1}{2}mV^2 t\right] + i\varphi_j(t)\right\} \quad (j=1, 2), \qquad (2.11)$$

where $\Phi_j(\xi)$ is a one-electron envelope function in terms of the running wave variable $\xi = (z - z_0 - Vt)/a$. Here $E_j$ is the eigen-energy, $\varphi_j(t)$ is the phase and $m = \hbar^2/(2Ja^2)$ is the effective mass of an electron. Note that in the absence of extra electrons in the lattice the wave velocity depends on the actual lattice compression, $\rho$, determined by Eq. (2.8), and may increase or decrease with amplitude, though for the Toda, Morse and Lennard-Jones potentials it is an always increasing function of the wave amplitude. Note also that for steady motions $V$ is constant.

In one-dimensional systems the deformational potential has at least one bound state, and the minimum of the energy of the system corresponds to the state when both electrons occupy the same level in the common potential well [12]. In the general case the maximum values of the electron wave functions are shifted along the lattice at some value, $la$, which is determined by the balance between the expected Coulomb repulsion between the electrons and their lattice-mediated attraction, similar to the case of binding of two extra electrons in a harmonic lattice [13]. Therefore, we can write

$$\Phi_j(\xi) = \Phi(\xi \pm l/2) f_j(l), \tag{2.12}$$

where we have considered that the functions $f_j(l)$ take into account the modulation of one-electron wave functions due to the Coulomb repulsion. If for localized states extending over several lattice sites the repulsion is weak enough then $f_j(l) \approx 1 + \varepsilon \phi_j(l)$ where $\varepsilon$ is a smallness parameter, $\varepsilon \ll 1$. In the lowest-order approximation with respect to $\varepsilon$ the maxima of both one-electron functions coincide at $\xi = 0$ (this is always possible by the appropriate choice of $z_0$), and, correspondingly, the two eigen-energies and eigen-functions are equal:

$$E_1 = E_2 \equiv E \quad \text{and} \quad \Phi_1(\xi) = \Phi_2(\xi) \equiv \Phi(\xi) . \tag{2.13}$$

For universality in the argument, let us introduce the following dimensionless parameters:

$$\lambda = -\frac{E}{J}, \quad \sigma = \frac{\chi a}{J}, \quad D = \frac{\chi a}{MV_{ac}^2} . \tag{2.14}$$

From Eqs (2.6) and (2.8) using (2.9) follows:

$$\frac{d^2 \Phi(\xi)}{d\xi^2} + \sigma \rho(\xi) \Phi(\xi) = \lambda \Phi(\xi) , \tag{2.15}$$

$$\Phi^2(\xi) = \frac{1}{2D} \frac{dF(\rho)}{d\rho} , \tag{2.16}$$

where

$$F(\rho) = U(\rho) - \frac{1}{2} s^2 \rho^2 = \frac{1}{4} \gamma \rho^2 (\rho^2 + 2\delta). \tag{2.17}$$

Here $s^2 = V^2 / V_{ac}^2$ and

$$\delta = \frac{1 - s^2}{\gamma} \tag{2.18}$$

is the dynamically modulated inverse anharmonic stiffness coefficient.

## 3. Localized solutions

From Eq. (2.15) we get

$$\left(\frac{d\Phi}{d\xi}\right)^2 = \lambda \Phi^2(\xi) - \sigma Q(\xi) \tag{3.1}$$

with

$$Q(\xi) = \int_{-\infty}^{\xi} \rho(x) d\Phi^2(x). \tag{3.2}$$

We search for localized solutions of Eqs. (2.15)-(2.16) which exponentially decay in space and attain some maximum values, which we denote by $\Phi_0$ and $\rho_0$, for the wave-function and corresponding deformation, respectively. From Eq. (3.1) we get the electron eigen-energy:

$$\lambda = \sigma \frac{Q(0)}{\Phi_0^2}. \tag{3.3}$$

From Eq. (2.16) we get

$$d\Phi^2(\xi) = \frac{1}{2D} d\left(\frac{dF}{d\rho}\right), \tag{3.4}$$

that substituted into Eq. (3.1) yields, after integration, the following equation:

$$F(\rho) = \frac{1}{2D} \int_0^{\rho(\xi)} \rho' d\left(\frac{dF}{d\rho'}\right) = \frac{1}{2D} \frac{dF}{d\rho} G(\rho), \tag{3.5}$$

where

$$G(\rho) = \rho - \frac{F(\rho)}{dF/d\rho} = \frac{\rho}{4} \frac{3\rho^2 + 2\delta}{\rho^2 + \delta}. \tag{3.6}$$

Differentiating Eq. (2.15) with respect to $\xi$, we get

$$\left(\frac{d\Phi(\xi)}{d\xi}\right)^2 = \frac{1}{8D} \frac{\left(d^2F/d\rho^2\right)^2}{dF/d\rho} \left(\frac{d\rho}{d\xi}\right)^2. \tag{3.7}$$

Then from Eq. (3.1) we have

$$\left(\frac{d\Phi(\xi)}{d\xi}\right)^2 = \frac{1}{2D} \frac{dF}{d\rho} (\lambda - \sigma G). \tag{3.8}$$

Eqs. (3.7) and (3.8) yield the equation for the lattice deformation:

$$\frac{d\rho}{d\xi} = \pm 2\sqrt{\sigma}\,\frac{dF/d\rho}{d^2F/d\rho^2}\sqrt{G(\rho_0)-G(\rho)}, \tag{3.9}$$

where we have taken into account the relation

$$\lambda = \sigma G(\rho_0), \tag{3.10}$$

which follows from Eq. (3.3). Indeed to obtain Eq. (3.9) we first get $\dfrac{d\rho}{d\xi}$ from Eq (3.7), take into account (3.2) and (3.4) and then integrate the result by parts:

$$Q(0) = \int_{-\infty}^{0} \rho(x)d\Phi^2(x) = \int_{-\infty}^{0} \rho(x)\frac{1}{2D}d\left(\frac{dF}{d\rho}\right) = \frac{1}{2D}\left[\frac{dF}{d\rho}\left(\rho - \frac{F}{dF/d\rho}\right)\right]_{\rho=\rho_0} = \Phi_0^2 G(\rho_0). \tag{3.11}$$

Then integrating Eq. (3.9), we get

$$\xi(\rho) = \pm\frac{1}{2\sqrt{\sigma}}\int_{\rho(\xi)}^{\rho_0}\frac{d^2F/d\rho^2}{dF/d\rho}\frac{1}{\sqrt{G(\rho_0)-G(\rho)}}d\rho. \tag{3.12}$$

The difference function in the kernel of Eq. (3.12) can be rewritten as

$$G(\rho_0) - G(\rho) = (\rho_0 - \rho)\Theta(\rho,\rho_0), \tag{3.13}$$

with

$$\Theta(\rho,\rho_0) = \frac{3\rho^2\rho_0^2 + \delta\left(3\rho^2 + \rho\rho_0 + 3\rho_0^2\right) + 2\delta^2}{4\left(\rho^2+\delta\right)\left(\rho_0^2+\delta\right)}. \tag{3.14}$$

Taking into account (3.13), Eq. (3.12) becomes

$$\xi(\rho) = \pm\frac{1}{\sqrt{2\sigma}}\int_{\rho(\xi)}^{\rho_0}\frac{K(\rho,\rho_0)}{\rho\sqrt{\rho_0-\rho}}d\rho, \tag{3.15}$$

where

$$K(\rho,\rho_0) = \frac{1}{\sqrt{2\Theta(\rho,\rho_0)}}\frac{3\rho^2+\delta}{\rho^2+\delta}. \tag{3.16}$$

Then in view of (3.14), the expression (3.16) shows a weak dependence of $K$ on its variables. Practically it remains in value very close to a constant equal to unity. The quantity $K(\rho, \rho_0)$ (3.16) is plotted in Fig. 1 for $\delta=0.5$. Note that $K$ varies 4% only, from 1 to 1.04. Only the part of the plot for $\rho < \rho_0$ has physical meaning since by definition $\rho_0$ defines the maximum value of $\rho$.

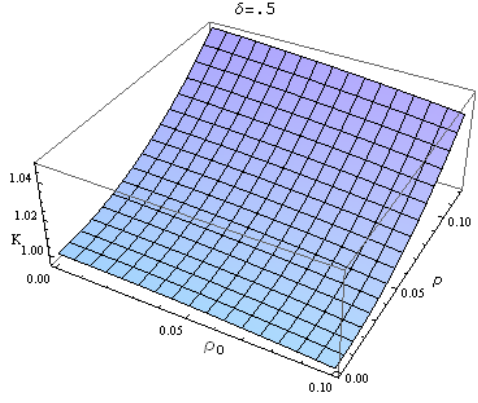

**FIGURE 1**. Dependence of $K(\rho, \rho_0)$ on the lattice deformation $\rho$, as function of the maximum $\rho_0$ for $\delta=0.5$.

Taking this into account, we can integrate Eq. (3.12), which gives

$$\rho(\xi) = \rho_0 Sech^2(\kappa\xi), \tag{3.17}$$

with

$$\kappa = \sqrt{\frac{\sigma\rho_0}{2}} \tag{3.18}$$

defining the inverse width spanned by the lattice deformation.

The maximum value of the lattice deformation can be found from the normalization condition (2.10), which takes the form

$$\frac{1}{D}\int_0^{\rho_0} \frac{dF}{d\rho}|d\xi(\rho)| = 1, \tag{3.19}$$

from which we get

$$\int_0^{\rho_0} \frac{d^2F/d\rho^2}{\sqrt{G(\rho_0)-G(\rho)}} d\rho = 2D\sqrt{\sigma}. \tag{3.20}$$

Substituting the relation (3.14) into Eq. (3.20) and using Eq. (2.17), we get

$$\int_0^{\rho_0} \frac{3\rho^2+\delta}{\sqrt{(\rho_0-\rho)\Theta(\rho,\rho_0)}} d\rho = \frac{2D}{\gamma}\sqrt{\sigma}. \tag{3.21}$$

Function $\Theta(\rho,\rho_0)$ shows quite a weak dependence on $\rho$ and can be approximated by its value at $\rho=\rho_0$:

$$\Theta(\rho,\rho_0) \approx \Theta(\rho_0,\rho_0) \equiv \vartheta(\rho_0) = \frac{3\rho_0^4 + 7\delta\rho_0^2 + 2\delta^2}{\rho_0^4 + 2\delta\rho_0^2 + \delta^2}. \qquad (3.22)$$

Figs. 2a and 2b illustrate the behavior of (3.14) and (3.22), respectively.

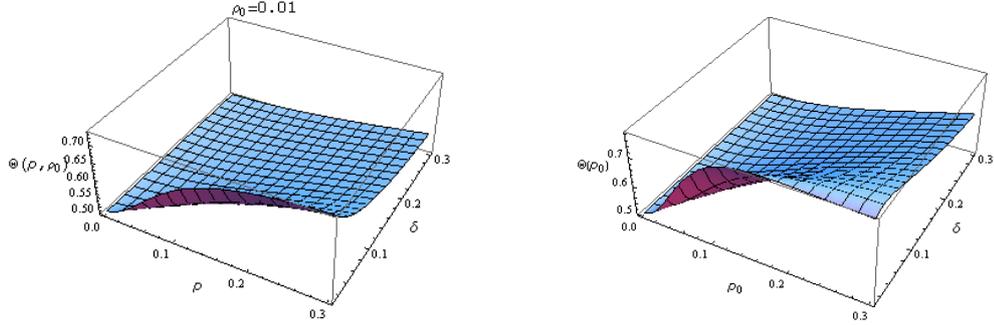

**FIGURE 2**. Left figure: $\Theta(\rho, \rho_0)$ *versus* lattice deformation, $\rho$, and the dynamically modulated inverse stiffness parameter, $\delta$, for $\rho_0=0.01$. Right figure: $\Theta(\rho_0, \delta)$ *versus* $\rho_0$ and $\delta$.

Using (3.22), we get from Eq. (3.21) the equation for the maximum value of deformation

$$\rho_0\left(\frac{8}{5}\rho_0^2 + \delta\right)^2 = \frac{\alpha^2}{4}\vartheta(\rho_0), \qquad (3.23)$$

where $\alpha$ is the dimensionless electron-lattice coupling constant:

$$\alpha^2 = \frac{4D^2\sigma}{\gamma^2}. \qquad (3.24)$$

The maximum value of the lattice deformation (3.23) as function of the dynamically modulated inverse anharmonic stiffness coefficient $\delta$, is depicted in Fig. 3 for two values of the electron-lattice coupling constant, $\alpha = 0.05$ and $0.2$. It appears that the faster is the bisolectron, the higher amplitude it has, determined by $\rho_0$, and the narrower its width is, $\Lambda = 2\pi a/\kappa$. Nevertheless, it does not shrink completely, its amplitude and width maintain finite values even when the bisolectron approaches the sound velocity, $V=V_{ac}$ ($s=1$ and $\delta=0$). This result is at variance to the case of the *harmonic* approximation in the lattice potential (2.8) [12,13]. For comparison, in Fig. 3 we also show the results for *cubic* anharmonicity (thin line) for the same values of $\alpha$. It appears that the stronger is the electron-lattice coupling constant, the bigger is the maximum value of the deformation. The *quartic* anharmonicity is dominant at small values of $\delta$ (2.18) which is the velocity rescaled inverse anharmonic stiffness coefficient.

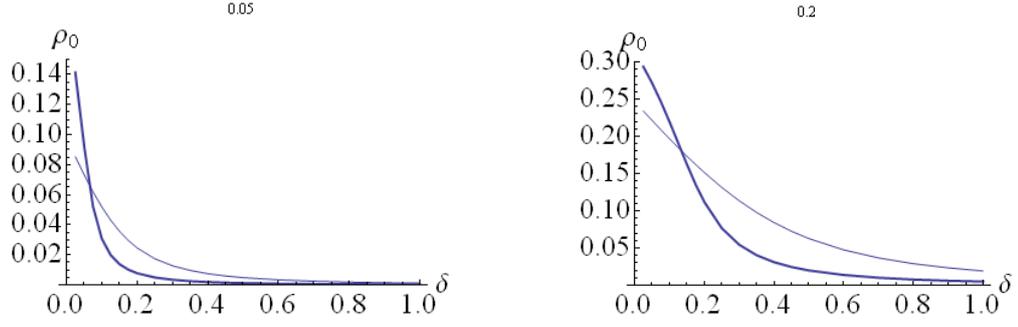

**FIGURE 3**. Maximum value of the lattice deformation $\rho_0$, as a function of the dynamically modulated inverse anharmonic stiffness coefficient $\delta$, in lattices with *quartic* (thick line) and *cubic* (thin line) anharmonicity for two values of electron-lattice coupling constant. Left figure: $\alpha=0.05$, right figure: $\alpha=0.2$.

Substituting the result (3.18) into Eq. (2.16) and using (2.17), we obtain the electron wave-function:

$$\Phi(\xi) = \sqrt{\frac{\rho_0}{2D}} Sech(\kappa\xi)\sqrt{1-s^2+\gamma\rho_0^2 Sech^4(\kappa\xi)}, \tag{3.25}$$

and the corresponding maximum of the wave function:

$$\Phi_0 = \sqrt{\frac{1}{2D}\left(\frac{dF}{d\rho}\right)\bigg|_{\rho=\rho_0}} = \sqrt{\frac{\gamma}{2D}\rho_0(\rho_0^2+\delta)}. \tag{3.26}$$

### 4. Energy of the bisoelectron and role of the Coulomb repulsion

Let us calculate the energy and momentum of the system. For (2.1) with the solutions (3.17), (3.25) (without Coulomb repulsion) we get

$$E_{tot}^{(bs)}(V) = mV^2 + E^{(bs)}(V) + W(V), \tag{4.1}$$

where the bisoelectron energy and energy of the lattice deformation are, respectively, given by the expressions:

$$E^{(bs)}(V) = -2\lambda J = -2DG(\rho_0)MV_{ac}^2, \tag{4.2}$$

$$W(V) = 2MV_{ac}^2\int_{-\infty}^{0}\!\!\left(F(\rho)+s^2\rho^2\right)\!d\xi = \frac{MV_{ac}^2}{\sqrt{\sigma}}\int_{0}^{\rho_0}\!\frac{d^2F/d\rho^2}{dF/d\rho}\frac{F(\rho)+s^2\rho^2}{\sqrt{G(\rho_0)-G(\rho)}}d\rho. \tag{4.3}$$

The total momentum of the system is

$$P(V) = \left(2m + M\int_{-\infty}^{\infty}\rho^2 d\xi\right)V = \left(2m + \frac{M}{\sqrt{\sigma}}\int_0^{\rho_0}\frac{d^2F/d\rho^2}{dF/d\rho}\frac{\rho^2}{\sqrt{G(\rho_0)-G(\rho)}}d\rho\right)V. \quad (4.4)$$

Substituting here the explicit expressions of $F$ given by (2.17) and its derivatives, we get the bisolectron energy, energy of the deformation and momentum of the system, respectively:

$$E^{(bs)}(V) = -\frac{1}{2}DMV_{ac}^2\rho_0\frac{3\rho_0^3 + 2\delta}{\rho_0^2 + \delta}. \quad (4.5)$$

$$W(V) \approx \frac{MV_{ac}^2}{2\sqrt{2\sigma}}\int_0^{\rho_0}K(\rho,\rho_0)\rho\frac{\gamma\rho^2 + 4\left(s^2 + \frac{1}{2}\delta\gamma\right)}{\sqrt{\rho_0-\rho}}d\rho \approx 8\frac{MV_{ac}^2}{\sqrt{2\sigma}}\rho_0^{3/2}\left[\frac{1}{3}\left(s^2 + \frac{1}{2}\delta\gamma\right) + \frac{2}{35}\gamma\rho_0^2\right]. \quad (4.6)$$

$$P(V) = \left(2m + M\sqrt{\frac{2}{\sigma}}\int_0^{\rho_0}K(\rho,\rho_0)\frac{\rho}{\sqrt{\rho_0-\rho}}d\rho\right)V \approx \left[2m + \frac{4}{3}M\sqrt{\frac{2}{\sigma}}\rho_0^{3/2}\right]V. \quad (4.7)$$

According to (2.11), the bisolectron moves along the lattice with constant velocity $V \leq V_{ac}$ and momentum

$$P = M_{eff}^{(bis)}V, \quad (4.8)$$

where $M_{eff}^{(bis)}$ is its *effective* mass, which, as it follows from the expression (4.7), is given by the relation

$$M_{eff}^{(bis)} \approx 2m + \frac{4}{3}M\sqrt{\frac{2J}{\chi a}}\rho_0^{3/2}. \quad (4.9)$$

It depends both on the velocity and lattice anharmonicity parameter through the corresponding dependence of the value of the maximum deformation, and, therefore, it attains the maximum value at $V = V_{ac}$. It is the essential feature of the anharmonic lattice that at the maximum value of the bisolectron velocity its *effective* mass and momentum are finite, hence, the *effective* mass approximation is valid in the whole interval of bisolectron velocities. Moreover, according to (4.1), the total energy of the system and the bisolectron energy are also finite at $V \to V_{ac}$. Accordingly, the lattice anharmonicity favors electron pairing in the whole interval of the bisolectron velocity.

Let us now take into account the Coulomb repulsion. We consider the stationary case, $V=0$. Then the solution (3.25), according to (2.12), takes the form

$$\Phi_i(\xi) \approx \sqrt{\frac{\rho_0}{2D}}Sech\left[\kappa\left(\xi \pm \frac{l}{2}\right)\right]\sqrt{1 + \gamma\rho_0^2 Sech^4\left[\kappa\left(\xi \pm \frac{l}{2}\right)\right]}, \qquad i=1,2. \quad (4.10)$$

where $l$ is the distance between the two maxima of the bisolectron function (2.12).

In Fig. 4 we show the one-electron wave functions (4.8) and the bisolectron wave-function which is proportional to the product of the latter two, and the charge distribution function (in electron charge units), i.e., $q(\xi) = \Phi_1^2(\xi) + \Phi_2^2(\xi)$ for two different values of *l*, which correspond to the distance between the two peaks in the bisolectron wave-function equal to two and three lattice sites, respectively. Note that the lattice deformation is proportional to $q(\xi)$, as it follows from Eq. (2.8) for the stationary case *V=0*. This is reminiscent of what is detected with the Scanning Tunneling Microscopy (STM). In related publications [5, 14-16] we have made use of this methodology for tracking soliton excitations both in 1D and 2D lattices.

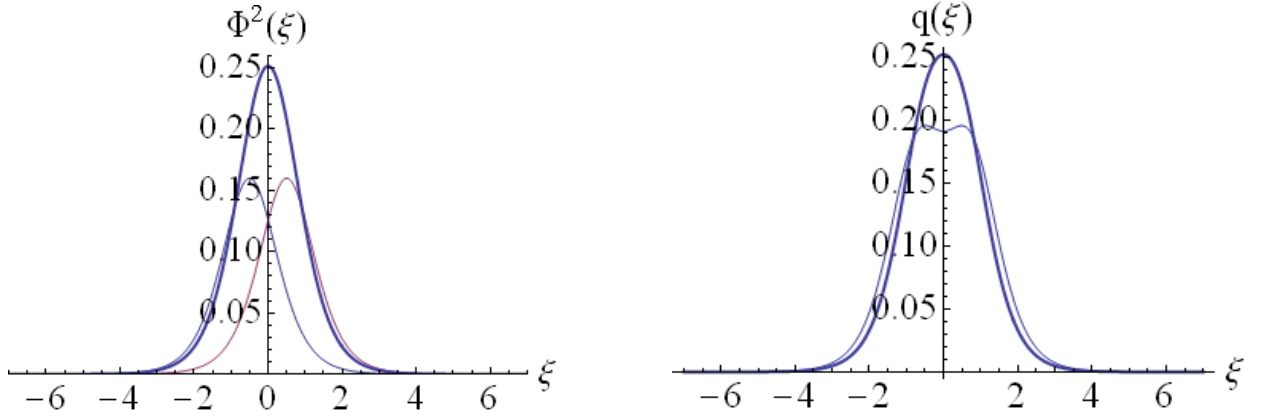

**FIGURE 4**. Left figure: one-electron wave functions (thin lines) and bisolectron wave function (thick line) with account of Coulomb repulsion, for $\kappa=1$ and $l=1$. Right figure: charge distribution over the lattice sites in the bisolectron state for $\kappa=1$ for $l=1$ (thick line) and $l=1.5$ (thin line).

As the bisolectron spans few lattice sites, we can use the approximation

$$E_c \approx \frac{e^2}{4\pi\varepsilon_0 la}, \qquad (4.11)$$

where *e* is the electron charge and $\varepsilon_0$ is the dielectric constant. Then to elucidate the role of the Coulomb repulsion we can approximate (4.10) by

$$\Phi_i(\xi) \approx \sqrt{\frac{\rho_0}{2D}} Sech\left[\kappa\left(\xi \pm \frac{l}{2}\right)\right]. \qquad (4.12)$$

Comparison of plots of functions (4.10) and (4.12) and their corresponding charge distributions shows that for physically relevant parameter values (not too high values of $\kappa$ to allow the wave functions spanning several lattice sites) the difference is negligibly small.

Substituting now (4.12) into the Hamiltonian (2.1) and expanding the wave functions and corresponding lattice deformation with respect to *l*, we obtain after integration the total energy of the system including the Coulomb repulsion (4.11):

$$E^{(bs)} = 2E_0 + \frac{2}{3}J\kappa^2\frac{\rho_0}{D} - \frac{4}{3}\frac{\chi a\rho_0^2}{\kappa D}\left(1 - l^2\kappa^2\right) + wa^2\rho_0^2\left[\frac{2}{3} + \frac{1}{2}\gamma\rho_0^2 - l^2\kappa^2\left(\frac{1}{3} + \frac{1}{2}\gamma\rho_0^2\right)\right] + \frac{e^2}{4\pi\varepsilon_0 la}.$$

(4.13)

Minimizing this expression with respect to the distance between the maxima of one-electron functions, *l*, we get the equilibrium distance

$$l_0 \approx \frac{1}{2}\left(\frac{e^2}{\pi\varepsilon_0 a\varsigma}\right)^{1/3} \quad (4.14)$$

with

$$\varsigma = \left[\frac{4\chi a\rho_0^2\kappa}{3D} - wa^2\rho_0^2\kappa^2\left(\frac{1}{3}+\frac{1}{2}\gamma\rho_0^2\right)\right]. \quad (4.15)$$

As $\rho_0 \ll 1$, the equilibrium distance between the wave function peaks can be approximated as

$$l_0 \approx \frac{1}{2}\left(\frac{3De^2}{4\pi\varepsilon_0\chi a^2\rho_0^2\kappa}\right)^{1/3}. \quad (4.16)$$

Then with the Coulomb repulsion between the electrons, $E_c$, the binding energy of the bisolectron is

$$E_{bind} = 2E^{(s)} - E^{(bs)} - E_c. \quad (4.17)$$

where $E^{(s)}$ is the energy of the system with a single added, excess electron as obtained in Refs. [11, 17]. Comparing $E^{(bs)}$ for $l=l_0$ with $E^{(s)}$ it results that $E_{bind} > 0$. Thus in spite of the stronger Coulomb repulsion in the bisolectron state, with spacing $l_0$ between the wave-function maxima, there is gain of energy due to pairing relative to the case of two separate electrons, placed at very large distance from each other.

## 5. Conclusions

It has been shown that the *quartic* anharmonicity in lattice interactions facilitates binding of two electrons in a singlet spin bisolectron state, spanning a few lattice sites. Such a bisolectron can move along the lattice with finite *effective* mass (4.9) and constant velocity. Moreover, the lattice anharmonicity limits bisolectron energy and momentum up to the sound velocity in the lattice whereas bisolitons in harmonic lattices are unstable at velocities higher than some critical value, $V \geq V_{cr}$, below the sound velocity, $V_{cr} < V_{ac}$ [13].

Due to the Coulomb repulsion between the electrons the two-electron wave function may have a one-hump envelope at not too strong repulsion or a two-hump envelope at high enough repulsion strengths. In the latter case the distance between the humps is determined by the balance between the gain of energy due to the coupling of electrons with the lattice and the lowering of energy due to the Coulomb repulsion. In systems with moderate electron-lattice interaction the bisolectron wave function is spread over few lattice sites and the distance between the two peaks of the bisolectron wave function is 2-3 lattice sites or more. In such a state the energy of the bisolectron is lower than the energy of the state with two separate, independent solectrons, and even with account of the Coulomb repulsion the bisolectron binding energy (4.15) is positive. Thus the bisolectron state (3.25) is indeed a bound state of two added, excess electrons with antiparallel spins in the common potential well determined by the soliton-like lattice deformation (3.17). The results here reported complement and confirm earlier fragmentary results obtained by computer simulations using the Gaussian approximation to the soliton excitation [18] and the harmonic and Morse potentials [19-21].

In view of the above it appears worth exploring the relationship of our bisolectron with Alexandrov's bipolaron [22]. This prediction was a finite small bipolaron bandwidth and a high critical superconducting temperature in the cross-over region from the BCS to bipolaronic superconductivity. What the bisolectron and a multipolaron and multisolectron dynamics share in common supporting the mentioned prediction?

Finally, let us give some examples of systems, in which bisolectrons can exist. The crucial reason for their existence is the validity of the adiabatic approximation used in deriving the equations of motion in the form (2.6)-(2.7). According to [23, 24], this approximation is valid for quasi-one-dimensional systems, whose dimesionless electron-phonon coupling constant $g = \frac{\chi^2}{2Jw}$ and non-adiabaticity parameter $\gamma_{non-ad} = \frac{\hbar V_{ac}}{2Ja}$ take values within certain intervals, which distinguish the regime of spontaneous electron localization as the crossover between the regimes of the formation of *small* polaron and almost free electron, respectively. To systems with such values of the electron-phonon coupling constant and non-adiabaticity parameter belong polydiacetylene [25-27], conducting platinum chain compounds such as RbCP(FHF), RbCP(DSH), CsCP(FHF), CsCP(Cl), KCP(Cl), CsCP($N_3$), KCP(Br), PbCP(Cl), GCP(Cl), ACP(Cl) [28]. There is also experimental evidence of the formation of *large* polarons in high-temperature superconducting cuprates [29-34] and salts of transition metals, such as PbSe, PbTe, PbS [35-38], at not too high values of the corresponding doping. Hence, we expect formation of bisolectrons in all the above cited materials. We plan to explore this issue in the future studies. The relation between the superconducting transition and the doping level in high-Tc superconductors points to on the close relation of this phenomenon with the electron-phonon coupling strength.

**ACKNOWLEDGEMENTS** The authors acknowledge enlightening discussions with Prof. R. Miranda and useful correspondence with Prof. S. Shastry. They are also grateful to three anonymous referees for enlightening remarks and suggestions that helped significantly improving the presentation of our results. This research was supported by the Spanish Ministerio de Ciencia e Innovacion under grant EXPLORA-FIS2009-06585-E. L. Brizhik also acknowledges partial support from the Fundamental Research Grant of the National Academy of Sciences of Ukraine.

**REFERENCES**

1. Velarde, M. G.; Brizhik, L.; Chetverikov, A. P.; Cruzeiro, L.; Ebeling, W.; Röpke, G. Int. J. Quantum Chem. 2011, 111 (to appear).
2. Christov, C. I.; Velarde, M. G. Int. J. Bifurcation Chaos 1994, 4, 1095.
3. Christov, C. I.; Maugin, G. A.; Velarde, M. G. Phys. Rev. E 1996, 54, 3621.
4. Nekorkin, V. I.; Velarde, M. G. Synergetic Phenomena in Active Lattices. Patterns, Waves, Solitons, Chaos; Springer-Verlag: Berlin, 2002, ch. 1.
5. Chetverikov, A. P.; Ebeling, W.; Velarde, M. G. Eur. Phys. J. B 2009, 70, 217.
6. Velarde, M. G. J. Computat. Applied Maths. 2010, 233, 1432.
7. Cantu Ros, O. G.; Cruzeiro, L.; Velarde, M. G.; Ebeling, W. Eur. Phys. J. B 2009, 80, 545.
8. Landau, L. D. Phys. Z. Sowjetunion, 1933, 3, 664.
9. Pekar, S. I. Untersuchungen uber die Elektronentheorie; Akademie Verlag: Berlin, 1954.
10. Fröhlich, H. Adv. Phys. 1954, 3, 325.
11. Davydov, A. S. Solitons in Molecular Systems, 2$^{nd}$ ed.; Reidel: Dordrecht, 1991.
12. Brizhik, L. S.; Eremko, A. A. Physica D 1995, 81, 295.
13. Brizhik, L. S.; Davydov, A. S. J. Low Temp. Phys. 1984, 10, 748.
14. Chetverikov, A. P.; Ebeling, W.; Velarde, M. G. Condens. Matter Phys. 2009, 12, 633.
15. Chetverikov, A. P.; Ebeling, W.; Velarde, M. G. Eur. Phys. J. B 2011, 80, 137.
16. Chetverikov, A. P.; Ebeling, W.; Velarde, M. G. Wave Motion 2011 (to appear).
17. Davydov, A. S.; Zolotaryuk, A. V. Phys. Lett. A 1983, 94, 49.
18. Velarde, M. G.; Neissner, C. Int. J. Bifurcation Chaos 2008, 18, 885.
19. Cruzeiro, L.; Eilbeck, J. C.; Marín, J. L.; Russell, F. M. Eur. Phys. J. B 2004, 42, 95.
20. Hennig, D.; Velarde, M. G.; Ebeling, W.; Chetverikov, A. P. Phys. Rev. E 2008, 78, 066606.
21. Velarde, M. G.; Ebeling, W.; Chetverikov, A. P. Int. J. Bifurcation Chaos 2011, 21,1595.
22. Alexandrov, A. S., Ed. Polarons in Advanced Materials; Springer: Dordrecht, 2007, pp. 257-310 (and references therein).
23. Brizhik, L. S.; Eremko, A. A.; La Magna, A.; Pucci. R. Phys. Lett. A, 1995, 205, 90.
24. Brizhik, L. S.; Eremko, A. A. Z. Phys. B, 1997, 104, 771.
25. Wilson, E. G. J: Phys. C, 1983, 16, 6739.
26. Donovan, K. J.; Wilson, E. G. Phil. Mag. B 1981, 44, 9.
27. Gogolin, A. A. Pis'ma ZhETP, 1986, 43, 395.
28. Carneiro, K. Properties of conducting platinum chain compounds, in: Electronic Properties of Inorganic Quasi-One- Dimensional Compounds; Monceau, P. Ed.; D. Reidel: Dordrecht, 1985; Part II, p. 1.
29. Reznik, D.; Pintschovius, L.; Ito, M.; Iikubo, S.; Sato, M.; Goka, H.; Fujita, M.; Yamada, K.; Gu, G. D. Nature 2006, 440, 1170.

30. Falter, C.; Hoffmann. G. A. Phys. Rev. B 2001, 64, 054516.
31. Bohnen, K.-P.; Heid, R.; Krauss, M. Europhys. Lett. 2003, 64, 104.
32. McQueeney, R. J.; Petrov, Y.; Egami, T.; Yethiraj, M.; Shirane, G.; Endoh, Y. Phys. Rev. Lett. 1999, 82, 628.
33. Devereaux, T. P.; Cuk, T.; Shen, Z.-X.; Nagaosa, N. Phys. Rev. Lett. 2004, 93, 117004.
34. Chung, J.-H.; Egami, T.; McQueeney, R. J.; Yethiraj, M.; Arai, M.; Yokoo, T.; Petrov, Y.; Mook, H. A.; Endoh, Y.; Tajima, S.; Frost, C.; Dogan, F. Phys. Rev. B 2003, 67, 014517.
35. Streetman, B. G.; Sanjay, B. Solid State Electronic Devices, 5th ed.; New Jersey: Prentice Hall, 2000, p. 524.
36. Zhang, Y.; Ke, X.; Chen, C; Kent, P.C. Phys. Rev. B 2009, 80, 024303.
37. Madelung, O.; Rössler, U.; Schultz, M., Eds. Lead selenide (PbSe) crystal structure, lattice parameters, thermal expansion; V. 41C: Non-Tetrahedrally Bonded Elements and Binary Compounds I. Springer Materials–The Landolt-Börnstein Database (http://www.springermaterials.com) DOI:10.1007/10681727_903
38. Androulakis; J.; Lee, Y.; Todorov, I.; Chung, D.-Y.; Kanatzidis, M. Phys. Rev. B 2011, 83, 195209.